\documentclass[%
 reprint,
superscriptaddress,
 amsmath,amssymb,
 aps,
prl,
]{revtex4-2}

\usepackage{graphicx}
\usepackage{graphicx}
\usepackage{dcolumn}
\usepackage{bm}
\usepackage{verbatim}
\usepackage[dvipsnames]{xcolor}
\usepackage{hyperref}
\usepackage[mathlines]{lineno}

\usepackage{ulem}
\begin{document}

\preprint{APS/123-QED}

\title{Effect of dispersion on indistinguishability between single-photon wave-packets}
\author{Yun-Ru Fan}
\affiliation{Institute of Fundamental and Frontier Sciences, University of Electronic Science and Technology of China, Chengdu 610054, China}
\author{Chen-Zhi Yuan}
\email{c.z.yuan@uestc.edu.cn}
\affiliation{Institute of Fundamental and Frontier Sciences, University of Electronic Science and Technology of China, Chengdu 610054, China}
\author{Rui-Ming Zhang}
\affiliation{Institute of Fundamental and Frontier Sciences, University of Electronic Science and Technology of China, Chengdu 610054, China}
\author{Si Shen}
\affiliation{Institute of Fundamental and Frontier Sciences, University of Electronic Science and Technology of China, Chengdu 610054, China}
\author{Peng Wu}
\affiliation{Institute of Fundamental and Frontier Sciences, University of Electronic Science and Technology of China, Chengdu 610054, China}
\author{He-Qing Wang}
\affiliation{Shanghai Institute of Microsystem and Information Technology, Chinese Academy of Sciences, Shanghai 200050, China}
\author{Hao Li}
\affiliation{Shanghai Institute of Microsystem and Information Technology, Chinese Academy of Sciences, Shanghai 200050, China}
\author{Guang-Wei Deng}
\affiliation{Institute of Fundamental and Frontier Sciences, University of Electronic Science and Technology of China, Chengdu 610054, China}
\author{Hai-Zhi Song}
\affiliation{Institute of Fundamental and Frontier Sciences, University of Electronic Science and Technology of China, Chengdu 610054, China}
\affiliation{Southwest Institute of Technical Physics, Chengdu 610041, China}
\author{Li-Xing You}
\affiliation{Shanghai Institute of Microsystem and Information Technology, Chinese Academy of Sciences, Shanghai 200050, China}
\author{Zhen Wang}
\affiliation{Shanghai Institute of Microsystem and Information Technology, Chinese Academy of Sciences, Shanghai 200050, China}
\author{You Wang}
\email{youwang\_2007@aliyun.com}
\affiliation{Institute of Fundamental and Frontier Sciences, University of Electronic Science and Technology of China, Chengdu 610054, China}
\affiliation{Southwest Institute of Technical Physics, Chengdu 610041, China}
\author{Guang-Can Guo}
\affiliation{Institute of Fundamental and Frontier Sciences, University of Electronic Science and Technology of China, Chengdu 610054, China}
\affiliation{CAS Key Laboratory of Quantum Information, University of Science and Technology of China, Hefei 230026, China}
\author{Qiang Zhou}
\email{zhouqiang@uestc.edu.cn}
\affiliation{Institute of Fundamental and Frontier Sciences, University of Electronic Science and Technology of China, Chengdu 610054, China}
\affiliation{CAS Key Laboratory of Quantum Information, University of Science and Technology of China, Hefei 230026, China}


\begin{abstract}
With propagating through a dispersive medium, the temporal-spectral profile of laser pulses should be inevitably modified. Although such dispersion effect has been well studied in classical optics, its effect on a single-photon wave-packet, i.e., the matter wave of a single-photon, has not yet been entirely revealed. In this paper, we investigate the effect of dispersion on indistinguishability of single-photon wave-packets through the Hong-Ou-Mandel (HOM) interference. By dispersively manipulating two indistinguishable single-photon wave-packets before interfering with each other, we observe that the difference of the second-order dispersion between two optical paths of the HOM interferometer can be mapped to the interference curve, indicating that (1) with the same amount of dispersion effect in both paths, the HOM interference curve must be only determined by the intrinsic indistinguishability between the wave-packets, i.e., dispersion cancellation due to the indistinguishability between Feynman paths; (2) unbalanced dispersion effect in two paths cannot be cancelled and will broaden the interference curve thus providing a way to measure the second-order dispersion coefficient. Our results suggest a more comprehensive understanding of the single-photon wave-packet and pave ways to explore further applications of the HOM interference.
\end{abstract}
\maketitle

Matter wave theory describes the wave property of physical objects \cite{Broglie1924}, which gives a comprehensive understanding of the wave-particle duality of quantum objects, for instance single-photons \cite{Dirac1930}. Similar to procedures for analyzing its classical counterpart, i.e., the electromagnetic wave, properties of the matter wave of single-photons have been investigated through interference\cite{Shields2009, Walmsley2017}. Exciting demonstrations, which certify the genuine quantum nature of single-photons, have been realized \cite{Guo2012, Peruzzo2012, Tanzilli2012}. The indistinguishability of single-photon wave-packet plays an important role in quantum information, which has been characterized through Hong-Ou-Mandel (HOM) interference in different degrees of freedom of single-photon wave-packets, such as in spatial mode\cite{Walborn2003}, temporal mode \cite{Ou2006}, polarization mode \cite{Moschandreou2018}, spectral mode \cite{Jin2015, Kobayashi2016, luo2019}, and orbital angular momentum mode \cite{Faccio2018}. Moreover, the HOM interference is also applied to  quantum communications with dispersive quantum channels, such as quantum teleportation \cite{Valivarthi2016, Sun2016} and measurement device independent quantum key distribution \cite{Rubenok2013, Yin2016, Tang2014}. Generally, the dispersion distortion should take place after a single-photon wave-packet propagating through dispersive environment. For instance, the dispersion induced temporal-mode broadening has been investigated with single-photon wave-packets in  Ref.~[\onlinecite{Kim2008}]. Furthermore, the dispersion effect would also change the indistinguishability between single-photon wave-packets in the spectral-temporal profile, which has not yet been revealed so far.

In this Letter, we investigate the effect of dispersion on indistinguishability between single-photon wave-packets via the HOM interference and implement two proof-of-principle demonstrations with coherent single-photon wave-packets which are prepared by attenuating mode-locked laser pulses. Our analyses show that
the difference of dispersion effect between the propagation paths of the two single-photon wave-packets can be mapped into the HOM interference curve. In one case, the mode-locked laser pulses propagate through a dispersion module, and then are separated into two parts which are attenuated to single-photon levels before sending them to a balanced HOM interferometer. Our results show that the width of HOM interference curve is independent with the dispersion from the dispersion module,  i.e., dispersion cancellation  \cite{Franson1992, Chiao1992, Chiao1992-2} due to the indistinguishability between Feynman paths \cite{Shih1996, Shih1998}, thus restoring the original temporal width of the wave-packets. In another case, two attenuated coherent single-photon wave-packets are sent into an unbalanced HOM interferometer, i.e.,  different dispersion between optical interference paths. The unbalanced dispersion cannot be cancelled and will modify the HOM interference curve. By measuring the change of the interference curve, we can obtain the second-order dispersion coefficient of the unbalanced dispersion, thus providing a new method for measuring it.

\begin{figure*}
\includegraphics[width=12.9 cm]{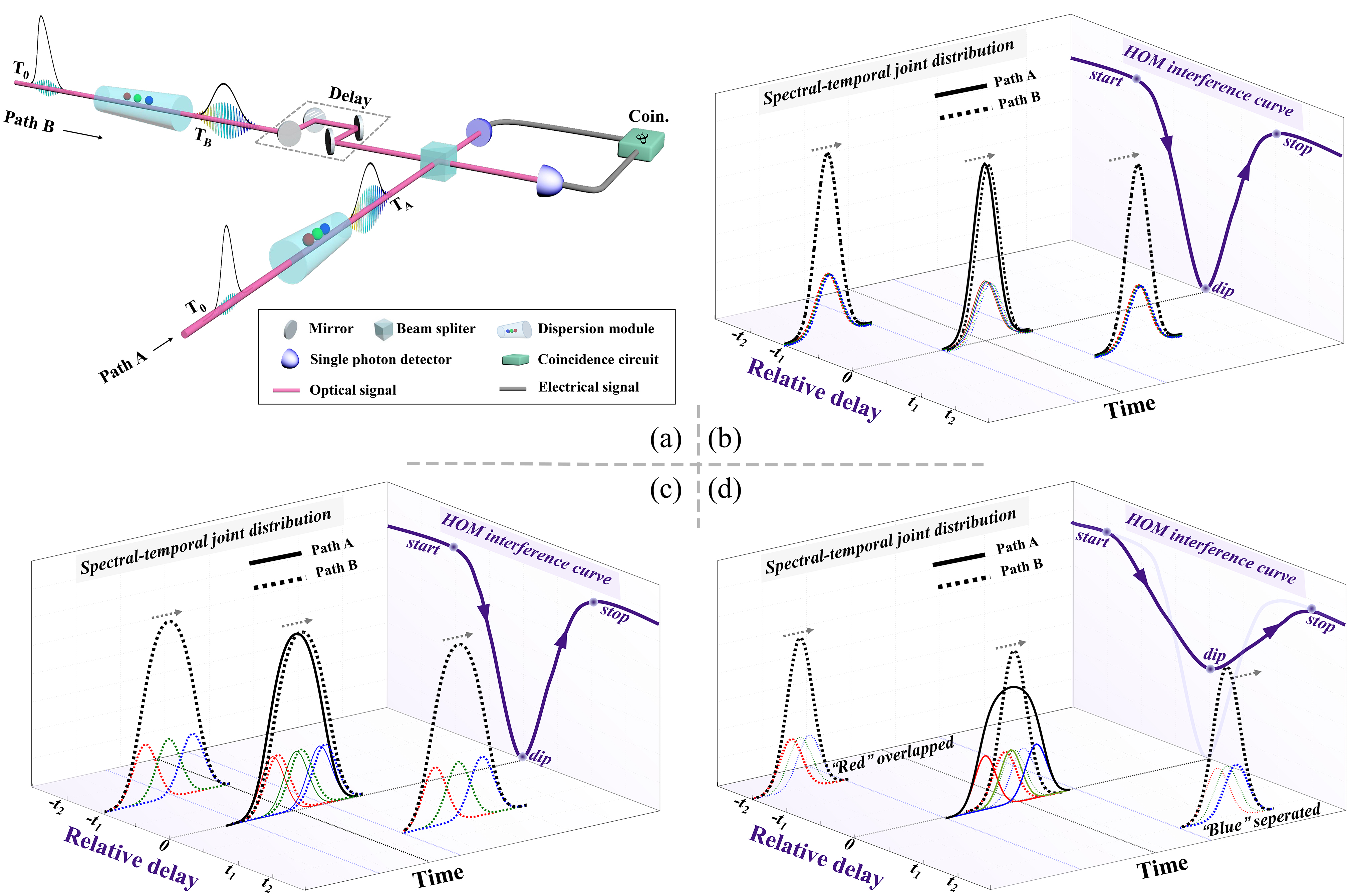}
\caption{\label{fig:fig1} Conceptual illustration of the HOM interference to reveal the dispersion effect on the indistinguishability between single-photon wave-packets. (a) HOM interferometer with dispersive manipulation modules. Two identical single-photon wave-packets are manipulated with dispersion modules along two optical paths, i.e., path A and path B, and then are sent into a HOM interferometer; (b) HOM interference curves without the second-order dispersion along two paths; (c) with the same second-order dispersion along two paths, i.e., balanced HOM interferometer; (d) with unbalanced second-order dispersions along two paths. To obtain the HOM interference curve, the travel time of the wave-packet in path A is fixed, and that in path B is varied and the time axis is in reference to the center of the pulse in path A. To guide eyes, envelopes of three sub-wave-packets are depicted with solid and dash lines in red, green, and blue respectively, and the black envelope covering the three sub-wave-packets are used to illustrate the widths of the wave-packets.}
\end{figure*}
Figure 1(a) gives the conceptual illustration of our proposed method. We consider two single-photon wave-packets with the pulse width of $T_0$, which propagate along path A and path B respectively. Two dispersion modules are utilized to manipulate the spectral-temporal profiles of two wave-packets. Then the wave-packets are input to a HOM interferometer, consisting of a 50:50 beam splitter (BS), two single photon detectors, and a coincidence circuit. To observe the HOM interference curve, an optical delay line is introduced in path B. The effect of dispersion on indistinguishability between two single-photon wave-packets is measured by comparing the HOM interference curves with different dispersive manipulations.

Assuming the two single-photon wave-packets in spatial modes A and B shown in Fig. 1 (a) are both in coherent state, we can present their initial quantum state as \cite{Blow1990}
\begin{equation}
\left|\psi_{0}\right\rangle_{A, B}=\left|\alpha_{A, B}(t)\right\rangle_{A}
\end{equation}
satisfying
\begin{equation}
a_{A, B}\left(t^{\prime}\right)\left|\alpha_{A, B}(t)\right\rangle_{A}=\alpha_{A, B}\left(t^{\prime}\right)\left|\alpha_{A, B}(t)\right\rangle_{A}
\end{equation}
and
\begin{equation}
\alpha_{A, B}(t)=\left(T_{0} \sqrt{\pi}\right)^{-1 / 2} e^{-\frac{t^{2}}{2 T_{0}^{2}}},
\end{equation}
where $a_{A, B}\left(t\right)$ is the annihilation operators of fields in modes A or B, and the temporal amplitude function $\alpha_{A, B}(t)$ is assumed to be Gaussian distribution.
The coincidence count per trial $P(\tau)$ without any dispersion can be expressed by (see more details in Supplemental Material),
\begin{equation}
\label{eq:1}
P(\tau)=1-\frac{1}{2} e^{-\frac{\tau^{2}}{2T_{0}^{2}}},
\end{equation}
where $\tau$ is the relative time delay between two optical paths. According to Eq.~{(\ref{eq:1})}, Fig. ~{\ref{fig:fig1}}(b) gives a conceptual HOM interference curve without dispersive manipulation and shows the cartoon process of how to obtain the HOM interference, in which the fixed single-photon wave-packet is indicated by solid lines (only shown with relative delay time being zero and omitted with other cases), while the moving one is indicated by dash lines.
 The HOM interference curve with a full width at half maximum (FWHM) which is $\sqrt{2}$ times of the FWHM of the two single-photon wave-packets can be obtained by scanning $\tau$ from $-t_1$ to $t_1$ in the axis of relative delay.

After the dispersive manipulation of a single-photon wave-packet, one may expect the interference curve to be changed due to the group-velocity time delay and the second-order dispersion effect, respectively. Ignoring the third and above order dispersion effects, we can obtain the coincidence count per trial $P^{'}(\tau)$, as given by (see more details in Supplemental Material)
\begin{equation}
\label{eq:2}
P^{'}(\tau)=1-\frac{T_{0}^{2}}{\sqrt{4 T_{0}^{4}+{\alpha}^{2}}} exp\left[{-\frac{2 \left(\tau-\delta\tau\right)^{2}}{4 T_{0}^{2}+\left({\alpha/T_{0}}\right)^{2}}}\right],
\end{equation}
where $\delta\tau=\beta_{1A} L_{A}-\beta_{1B} L_{B}$, $\alpha=\beta_{2 A} L_{A}-\beta_{2 B} L_{B}$ represent the difference of group-velocity time delay and the difference of the group-velocity dispersion propagating along two optical paths respectively, and $\beta_{1A}$ and $\beta_{1B}$ are group-velocity time delays, $\beta_{2A}$ and $\beta_{2B}$ are the group-velocity dispersions, $L_{A}$ and $L_{B}$ are the lengths of two dispersive optical paths.

The Eq.~{(\ref{eq:2})} gives the possible phenomena that would take place with dispersive manipulation. Such phenomena are further depicted in Figs.~{\ref{fig:fig1}}(c)-(d). In our cartoonish picture, the single-photon wave-packets are composed of three frequency components which are shown in red, green, and blue lines, respectively. Figure~{\ref{fig:fig1}}(c) indicates the situation that the two single-photon wave-packets experience the same amount of dispersion. Although for each wave-packet, different components in frequency are separated in time domain after the dispersive manipulation - temporal broadening, the HOM interference only occurs with components in the same color, i.e., with the relative  delay time from $-t_{1}$ to $t_{1}$ - as the same as the case shown in Fig.~{\ref{fig:fig1}}(b), which corresponds to the original width of wave-packets. On the other hand, the situation for two wave-packets experiencing different amount of dispersion is given in Fig. ~{\ref{fig:fig1}}(d). In this case, both of them are also broadened in time domain with different amounts, resulting in partially distinguishable. The HOM interference can still happen due to  the residual indistinguishability between them when the components in the same frequency overlapping, leading to a wider HOM interference curve with a smaller visibility. The corresponding changes of  the HOM interference curve are determined by the different amount of dispersion experienced by the two wave-packets, which is given by  Eq.~{(\ref{eq:3})},
\begin{equation}
\label{eq:3}
\alpha=T_{0} \sqrt{d^{2}/2 \ln 2-4 T_{0}^{2}},
\end{equation}
where $d$ is the FWHM of the HOM interference curve(details in Supplemental Material). Therefore, the different amount of dispersion can be obtained by measuring the width of the HOM interference curve.

Our proof-of-principle experimental setup is shown in Fig.~{\ref{fig:fig2}} (see more details in Supplemental Material). In the demonstrations, single-photon wave-packets are attenuated from mode-locked laser pulses with a mean photon number of 0.03 per pulse. The pulse width is $0.798\pm0.010$ ps, measured by a second-order autocorrelator (FEMTOCHROME, FR-103XL), and the period $T$ is 200 ns~\cite{Wu2019}. The dispersive manipulation is realized by using pieces of fiber as dispersion modules.
\begin{figure}
\includegraphics[width=8.6cm]{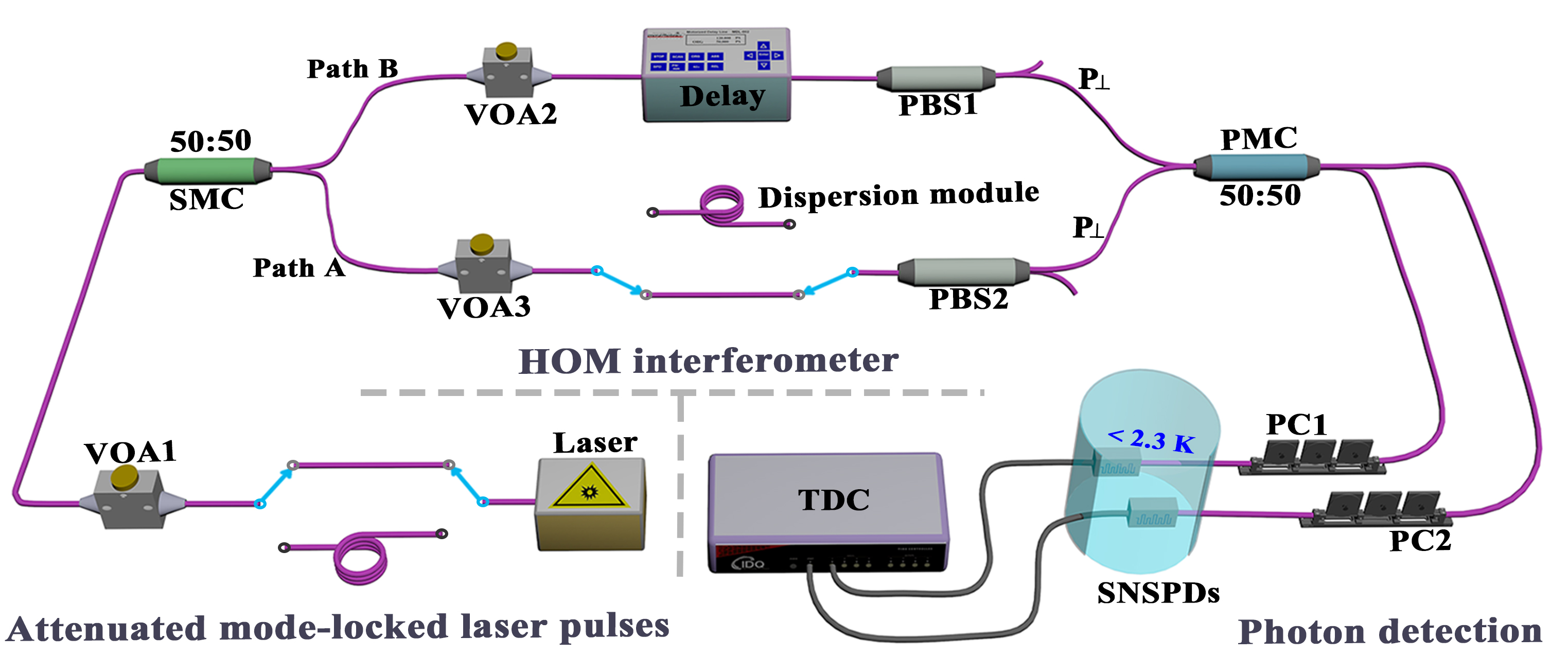}
\caption{\label{fig:fig2}Experimental setup. The setup consists of attenuated mode-locked laser pulses, HOM interferometer, and photon detection with a dispersion module. VOA: variable optical attenuator; SMC: single-mode fiber coupler; PBS: polarization beam splitter; PMC: polarization-maintaining fiber coupler; PC: polarization  controller; SNSPD: superconducting nanowire single photon detector; TDC: time to digital convertor.}
\end{figure}
We firstly measure the HOM interference curve at the output port of a mode-locked laser as shown in Fig.~{\ref{fig:fig2}}, which is used as a reference for our demonstrations. The HOM interference curve is obtained as shown in Fig.~{\ref{fig:fig3}}(a). The blue dots are the normalized experimental results and the solid lines are the Gaussian fitting curves with the Monte Carol method \cite{Molmer193}. From the fitted curves and  Eq.~{(\ref{eq:1})}, a FWHM HOM-dip of $1.035\pm0.008$ ps and a laser pulse width of $0.732\pm0.006$ ps can be obtained, respectively.

Next, we connect a dispersion module (50 km-long single-mode fiber spool, Yangtze Optical Fibre and Cable Co. LTD, G.652.D ULL) at the output port of our mode-locked laser. After propagating through this dispersive environment, the laser pulses would be broadened to about 3.9 ns with a second-order dispersion coefficient of 17.1 ps/(km$\cdot$nm). We employ a superconducting nanowire single photon detector (SNSPD, Photon Technology Co., P-CS-6)~\cite{Zhang2017} and a time to digital converter (TDC, ID Quantique, ID900) to measure the width of the broadened laser pulses after being attenuated to single-photon level. After the dispersive manipulation the measured pulse width is 4.1 ns - including a total jitter of about 270 ps (see more details in Fig. S1 in Supplemental Material). However, the width of the measured HOM interference curve is $1.032\pm0.009$ ps as shown in Fig.~{\ref{fig:fig3}}(b), which is as the same as that without the dispersive manipulation. This phenomenon would be explained as dispersion cancellation, i.e. the result of HOM interference is immune to the dispersion effect on individual wave-packet. The dispersion cancellation - including the dispersion related chirp effect \cite{Xiaoying2011, Xiaoying2015}-  has also been demonstrated with correlated two-photon pairs, in which the dispersion cancellation is attributed to the nonlocal property of entanglement. In our case, the dispersion cancellation phenomenon is related to the HOM interference based entangled N00N state generation \cite{Jakeman1990, Ou2006-2, Silberberg2010, Steinberg2014, Kok2016}, and can also be explained with Feynman diagram, i.e., we can not tell which Feynman path the single-photon wave-packet is from even with dispersive manipulation.
Therefore the HOM interference must recover the original width of the dispersively manipulated wave-packet, offering us a dispersion immune method to measure the intrinsic width of lase pulses. Although the HOM interference scheme has been used  to measure the width of ultrashort laser pulses in 1993~\cite{Matsuoka1993}, the dispersion influence on such a measurement has not yet been discussed, which is addressed in this work. It is worth noting that the pulse width from the HOM interference is $0.732\pm0.006$ ps, which is smaller than $0.798\pm0.010$ ps obtained from the second-order autocorrelation measurement~\cite{Giordmaine1966}. This would be caused by the dispersive broadening before the second-order autocorrelation measurement.

\begin{figure}
\includegraphics[width=8.6cm]{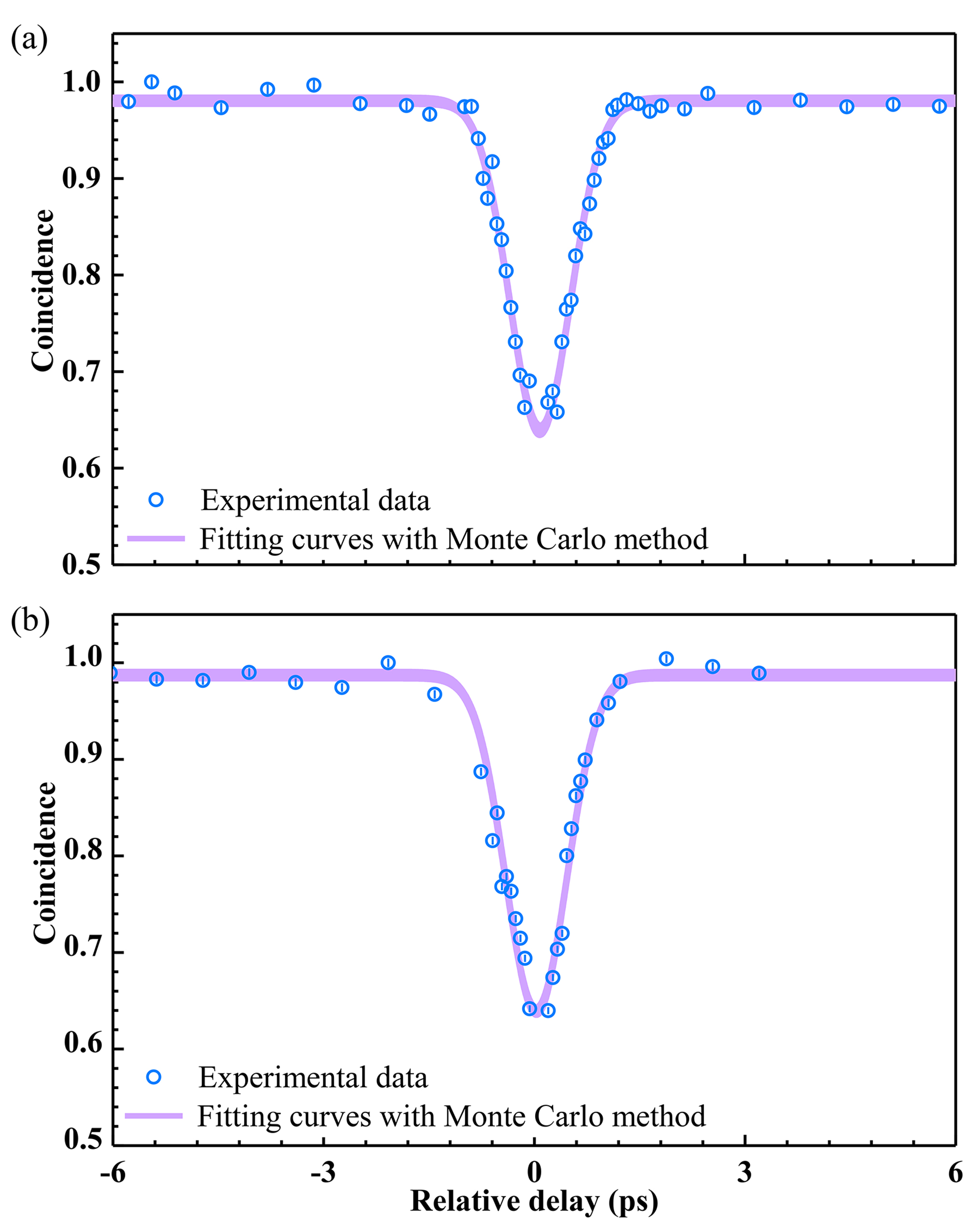}
\caption{\label{fig:fig3}HOM interference curves (a) without a dispersion module, and (b) with 50 km-long fiber as the dispersion module at the output of mode-locked laser, respectively. The blue dots are experimental results. The solid purple lines are 1000-time Monte Carlo fitting curves.}
\end{figure}

\begin{figure}
\includegraphics[width=8.6 cm]{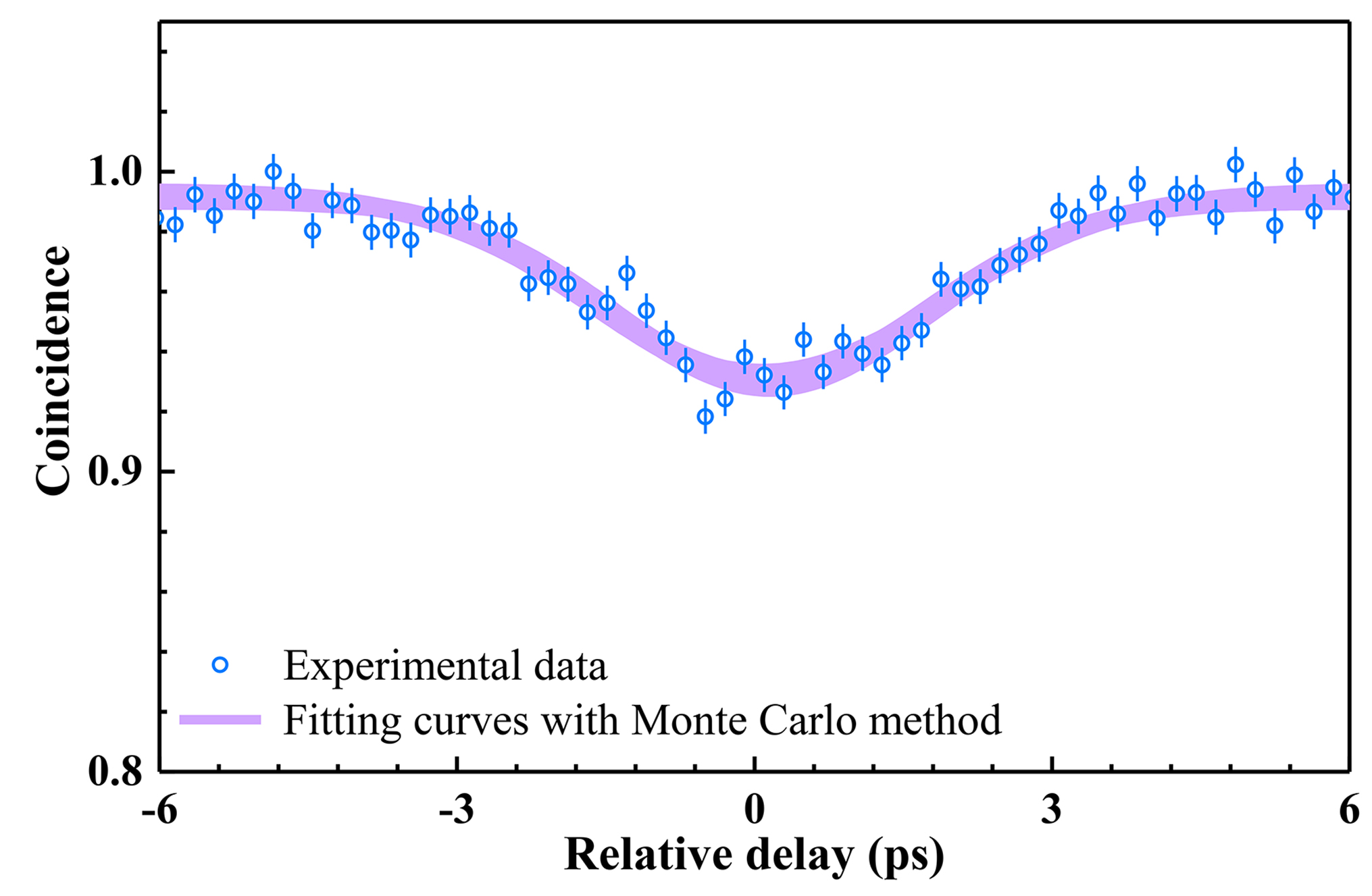}
\caption{\label{fig:fig4}HOM interference curves with a 80 m-long single-mode fiber inserted in one path, which correspond to two periods of mode-locked laser pulses.
}
\end{figure}

Then, we use another dispersion module (80 m-long single-mode fiber, SMF-28) to serve as the unknown optical material, which is inserted in one arm of the HOM interferometer as shown in Fig.~{\ref{fig:fig2}}. One may think that an extra optical path should be added in the other arm to balance transmission times in two arms. Fortunately, the extra optical path can be removed according to the result reported in Ref.~[\onlinecite{Tang2013}], i.e., the HOM interference can occur between periodically delayed mode-locked laser pulses. In our experiment, the HOM interference curves are measured between laser pulses with a two-period delay. The measured results are shown in Fig.~{\ref{fig:fig4}}. The FWHM of HOM interference curves is $4.123\pm0.124$ ps with the visibility of $0.062\pm0.01$, which is in consistence with our theoretical predictions.

According to Eq.~{(\ref{eq:1})}, the second-order dispersion coefficient of the inserted fiber is $15.04\pm0.48$ ps/(km$\cdot$nm), which is in consistence with the dispersion parameter from the manufacturer.
Several methods have been demonstrated to measure the second-order dispersion coefficient of optical media, for instance, the time of flight \cite{Cohen1977}, phase-shift \cite{Watanabe1976}, and interferometric methods \cite{Diels1996}. In the former two methods, the response time or time jitter of electronics limits the measurement precision \cite{Cohen1985}. As for the interferometric methods, the second-order dispersion coefficient is obtained from the first-order interference, which is phase sensitive. In our method, the second-order dispersion is measured by the HOM interference which  is insensitive in phase \cite{Mandel1989}. One may concern that the statistical properties of HOM method would influence the precision of our measurement. To address this, in our demonstration the measured results are not straightly extracted from the photon counting results but the Monte Carlo method fitted curves. The results are shown in Figs.~{\ref{fig:fig3}} and ~{\ref{fig:fig4}}, the solid purple lines indicate 1000-time Monte Carlo fitting curves. It can be seen that such a method
achieves a precision of 6 fs for the pulse width and a precision as high as $0.48$ ps/(km$\cdot$nm) for the second-order dispersion coefficient measurement.

In summary, we have investigated the effect of dispersion on indistinguishability between single-photon wave-packets via HOM interference. From the perspective of Feynman paths, the indistinguishability is independent of the same dispersive manipulation and only related to the unbalanced dispersion effect in two optical paths of the HOM interferometer. Our proof-of-principle demonstrations have proved that such a scheme can be used to measure the original pulse width of ultrashort laser pulses even after propagating through the dispersive environment, i.e., dispersion immune, and to measure the second-order dispersion coefficient of an unknown optical material. Our results show that the measurement precision can respectively reach 6 fs and 0.48 ps/(km$\cdot$nm) for the measurement of the pulse width and the second-order dispersion coefficient, suggesting that our schemes can be considered as practical methods. Our work provides a deeper understanding for single-photon wave-packets, i.e., matter waves of the single-photon, and opens up new applications for HOM interference.

\begin{acknowledgments}
The authors thank Professor Z. Y. Ou and Professor Xiaoying Li  for useful discussions. This work is partially supported by National Key Research and Development Program of China (Nos. 2018YFA0307400, 2017YFA0304000, 2018YFA0306102, 2017YFB0405100); National Natural Science Foundation of China (Nos. 61775025, 91836102, U19A2076, 61705033, 61405030, 61308041, 61704164, 62005039); China Postdoctoral Science Foundation (No. 2020M673178).
\end{acknowledgments}


\end{document}


\preprint{APS/123-QED}

\title{Effect of dispersion on indistinguishability between single-photon wave-packets}
\author{Yun-Ru Fan}
\affiliation{Institute of Fundamental and Frontier Sciences, University of Electronic Science and Technology of China, Chengdu 610054, China}
\author{Chen-Zhi Yuan}
\email{c.z.yuan@uestc.edu.cn}
\affiliation{Institute of Fundamental and Frontier Sciences, University of Electronic Science and Technology of China, Chengdu 610054, China}
\author{Rui-Ming Zhang}
\affiliation{Institute of Fundamental and Frontier Sciences, University of Electronic Science and Technology of China, Chengdu 610054, China}
\author{Si Shen}
\affiliation{Institute of Fundamental and Frontier Sciences, University of Electronic Science and Technology of China, Chengdu 610054, China}
\author{Peng Wu}
\affiliation{Institute of Fundamental and Frontier Sciences, University of Electronic Science and Technology of China, Chengdu 610054, China}
\author{He-Qing Wang}
\affiliation{Shanghai Institute of Microsystem and Information Technology, Chinese Academy of Sciences, Shanghai 200050, China}
\author{Hao Li}
\affiliation{Shanghai Institute of Microsystem and Information Technology, Chinese Academy of Sciences, Shanghai 200050, China}
\author{Guang-Wei Deng}
\affiliation{Institute of Fundamental and Frontier Sciences, University of Electronic Science and Technology of China, Chengdu 610054, China}
\author{Hai-Zhi Song}
\affiliation{Institute of Fundamental and Frontier Sciences, University of Electronic Science and Technology of China, Chengdu 610054, China}
\affiliation{Southwest Institute of Technical Physics, Chengdu 610041, China}
\author{Li-Xing You}
\affiliation{Shanghai Institute of Microsystem and Information Technology, Chinese Academy of Sciences, Shanghai 200050, China}
\author{Zhen Wang}
\affiliation{Shanghai Institute of Microsystem and Information Technology, Chinese Academy of Sciences, Shanghai 200050, China}
\author{You Wang}
\email{youwang\_2007@aliyun.com}
\affiliation{Institute of Fundamental and Frontier Sciences, University of Electronic Science and Technology of China, Chengdu 610054, China}
\affiliation{Southwest Institute of Technical Physics, Chengdu 610041, China}
\author{Guang-Can Guo}
\affiliation{Institute of Fundamental and Frontier Sciences, University of Electronic Science and Technology of China, Chengdu 610054, China}
\affiliation{CAS Key Laboratory of Quantum Information, University of Science and Technology of China, Hefei 230026, China}
\author{Qiang Zhou}
\email{zhouqiang@uestc.edu.cn}
\affiliation{Institute of Fundamental and Frontier Sciences, University of Electronic Science and Technology of China, Chengdu 610054, China}
\affiliation{CAS Key Laboratory of Quantum Information, University of Science and Technology of China, Hefei 230026, China}

\date{\today}

\maketitle

\section{\label{app:1}Theoretical scheme}
The Paths A and B in Fig. 1 (a) are defined as spatial modes A and B, respectively. Assuming the single-photon wave-packets in spatial modes A and B are both in coherent state, we can present their initial quantum state as \cite{Blow1990}
\begin{equation}
\label{eq:s1}
\left|\psi_{0}\right\rangle_{A, B}=\left|\alpha_{A, B}(t)\right\rangle,
\end{equation}
satisfying
\begin{equation}
\label{eq:s2}
a_{A, B}\left(t^{\prime}\right)\left|\alpha_{A, B}(t)\right\rangle=\alpha_{A, B}\left(t^{\prime}\right)\left|\alpha_{A, B}(t)\right\rangle,
\end{equation}
and
\begin{equation}
\label{eq:s3}
\int d t\left|\alpha_{A, B}(t)\right|^{2}=\left|\alpha_{A0, B0}\right|^{2},
\end{equation}
where $a_{A, B}\left(t^{\prime}\right)$ is the annihilation operators of fields in modes A or B, and $\left|\alpha_{A 0, B 0}\right|^{2}$ is the averaged photon number in each wave-packet. After passing through the dispersive mediums in the propagation pathways, the coherent state evolves from Eq.~{(\ref{eq:s1})} into
\begin{equation}
\label{eq:s4}
\left|\psi_{1}\right\rangle_{A, B}=\left|\alpha_{A 1, B 1}(t)\right\rangle,
\end{equation}
where
\begin{equation}
\label{eq:s5}
\alpha_{A 1, B 1}(t)=(2 \pi)^{-1 / 2} \int d \Omega_{A, B} \alpha_{A, B}\left(\Omega_{A, B}\right) e^{-i\left(\omega_{0}+\Omega_{A, B}\right) t+i \beta_{A, B}\left(\Omega_{A, B}\right) L_{A, B}},
\end{equation}
with $\alpha_{A, B}\left(\Omega_{A, B}\right)=(2 \pi)^{-1 / 2} \int d t \alpha_{A, B}(t) e^{i\left(\omega_{0}+\Omega_{A, B}\right) t}$ and the central angular frequency of the two wave-packets both assumed to be $ \omega_0$. In Eq.~{(\ref{eq:s5})}, $L_{A, B}$ and $\beta_{A, B}$ are the length of the dispersive medium on path A or B, and the corresponding propagation constant at angular frequency of $\omega_{0}+\Omega_{A, B}$. Before incident on the beam splitter in Fig. 1 (a), the wave-packet in mode B is delayed by $\tau$ and its quantum state turns into
\begin{equation}
\left|\psi_{2}\right\rangle_{B}=\left|\alpha_{B 1}(t+\tau)\right\rangle_{B}.
\end{equation}
\label{eq:s6}
Introducing spatial modes C and D corresponding to the two output ports of the beam splitter, we can obtain the annihilation operators $a_{C, D}(t)$ of fields in the two modes via the transforms
\begin{equation}
\label{eq:s7}
a_{C}(t)=(2)^{-1 / 2}\left[a_{A}(t)+a_{B}(t)\right],
\end{equation}
\begin{equation}
\label{eq:s8}
a_{D}(t)=(2)^{-1 / 2}\left[a_{A}(t)-a_{B}(t)\right].
\end{equation}
The second-order correlation between fields in modes C and D can be obtained as
\begin{equation}
\label{eq:s9}
G_{C, D}^{(2)}\left(t, t+\tau_{c}\right)={ }_{B}\left\langle\left.\psi_{2}\right|_{A}\left\langle\psi_{1}\left|a_{C}^{\dagger}(t) a_{D}^{\dagger}\left(t+\tau_{c}\right) a_{D}\left(t+\tau_{c}\right) a_{C}(t)\right| \psi_{1}\right\rangle_{A} \mid \psi_{2}\right\rangle_{B},
\end{equation}
where $\tau_{c}$ is the time delay introduced in calculating the second-order correlation function. Then, we can calculate the coincidence counts between the photons detected in modes C and D as
\begin{equation}
\label{eq:s10}
n_{C D}(\tau)=\eta_{C} \eta_{D} \int_{T_{1}}^{T_{2}} d t \int_{-\Delta \tau_{c 0} / 2}^{\Delta \tau_{c 0} / 2} d \tau_{c} G_{C, D}^{(2)}\left(t, t+\tau_{c}\right),
\end{equation}
where $\eta_{C, D}$ summarize the collection and detection efficiencies of the photons;   $T_{1}$ and $T_{2}$  are the start and stop time of the coincidence counting measurement; $\Delta \tau_{c 0}$ is the width of the coincidence counts window.
After substituting Eqs.~{(\ref{eq:s7})}-{(\ref{eq:s9})}
into Eq.~{(\ref{eq:s10})}, $n_{C D}(\tau)$ constitute of eight terms possible to be nonzero. We can list them as:

(i)
\begin{equation}
\label{eq:s11}
\begin{aligned}
n_{C D, 1}(\tau)=&\frac{\eta_{C} \eta_{D}}{4} \int_{T_{1}}^{T_{2}} d t \int_{-\Delta \tau_{c 0} / 2}^{\Delta \tau_{c 0} / 2} d \tau_{c}\left\langle\left.\psi_{1}\right|_{A}\left\langle\left.\psi_{2}\right|_{B} a_{A}^{\dagger}(t) a_{A}^{\dagger}\left(t+\tau_{c}\right) a_{A}\left(t+\tau_{c}\right) a_{A}(t) \mid \psi_{1}\right\rangle_{A} \mid \psi_{2}\right\rangle_{B} \\
=&\frac{\eta_{C} \eta_{D}}{4} \int_{T_{1}}^{T_{2}} d t \int_{-\Delta \tau_{c 0} / 2}^{\Delta \tau_{c 0} / 2} d \tau_{c}\left|\alpha_{A 1}(t)\right|^{2}\left|\alpha_{A 1}\left(t+\tau_{c}\right)\right|^{2} \approx \frac{\eta_{C} \eta_{D}}{4}\left|\alpha_{A0}\right|^{4}.
\end{aligned}
\end{equation}

 (ii)
\begin{equation}
\label{eq:s12}
\begin{aligned}
n_{C D, 2}(\tau) =&\frac{\eta_{C} \eta_{D}}{4} \int_{T_{1}}^{T_{2}} d t \int_{-\Delta \tau_{c 0} / 2}^{\Delta \tau_{c 0} / 2} d \tau_{c}\left\langle\left.\psi_{1}\right|_{A}\left\langle\left.\psi_{2}\right|_{B} a_{B}^{\dagger}(t) a_{B}^{\dagger}\left(t+\tau_{c}\right) a_{B}\left(t+\tau_{c}\right) a_{B}(t) \mid \psi_{1}\right\rangle_{A} \mid \psi_{2}\right\rangle_{B} \\
=&\frac{\eta_{C} \eta_{D}}{4} \int_{T_{1}}^{T_{2}} d t \int_{-\Delta \tau_{c 0} / 2}^{\Delta \tau_{c 0} / 2} d \tau_{c}\left|\alpha_{B 1}(t)\right|^{2}\left|\alpha_{B 1}\left(t+\tau_{c}\right)\right|^{2} \approx \frac{\eta_{C} \eta_{D}}{4}\left|\alpha_{B0}\right|^{4}.
\end{aligned}
\end{equation}

At the last step in Eq.~{(\ref{eq:s12})}, we assume that $T_{2}-T_{1}$ and $\Delta \tau_{c 0}$ is remarkably larger than the temporal width of the dispersion-broadened wave-packets, and this is easily satisfied in our experiment. It is obvious that $n_{C D, 1}(\tau)$ and $ n_{C D, 2}(\tau)$ originate from the second-order auto-correlation of the two wave-packets, respectively. In fact, in a Hanbury-Brown-Twiss experiment, the normalized second-order auto-correlation function of the two wave-packets is
\begin{equation}
\label{eq:s13}
g_{A, B}^{(2)}\left(\tau_{c}\right)=\frac{\int_{T_{1}}^{T_{2}} d t \int_{\tau_{c}-\Delta \tau_{c 0} / 2}^{\tau_{c}+\Delta \tau_{c 0} / 2} d \tau_{c A, B}^{\prime}\left\langle\psi_{1,2}\left|a_{A}^{\dagger}(t) a_{A}^{\dagger}\left(t+\tau_{c}^{\prime}\right) a_{A}\left(t+\tau_{c}^{\prime}\right) a_{A}(t)\right| \psi_{1,2}\right\rangle_{A, B}}{\left(\int_{T_{1}}^{T_{2}} d t_{A, B}\left\langle\psi_{1,2}\left|a_{A}^{\dagger}(t) a_{A}(t)\right| \psi_{1,2}\right\rangle_{A, B}\right)^{2}}.
\end{equation}
Thus,a general form of $n_{C D, 1}(\tau)$ and $ n_{C D, 2}(\tau)$ are

\begin{equation}
\begin{aligned}
\label{eq:s14-1}
n_{C D, 1}(\tau)=\frac{\eta_{C} \eta_{D}}{4} g_{A}^{(2)}(0)\left(\int_{T_{1}}^{T_{2}} d t_{A}\left\langle\psi_{1}\left|a_{A}^{\dagger}(t) a_{A}(t)\right| \psi_{1}\right\rangle_{A}\right)^{2}
\end{aligned}
\end{equation}
and
\begin{equation}
\begin{aligned}
\label{eq:s14-2}
n_{C D, 2}(\tau)=\frac{\eta_{C} \eta_{D}}{4} g_{B}^{(2)}(0)\left(\int_{T_{1}}^{T_{2}} d t_{B}\left\langle\psi_{2}\left|a_{B}^{\dagger}(t) a_{B}(t)\right| \psi_{2}\right\rangle_{B}\right)^{2}.
\end{aligned}
\end{equation}
Since we assume the single-photon wave-packets are in coherent state, $g_{A, B}^{(2)}(0)=1$ and therefore the results in Eq.~{(\ref{eq:s11})} (Eq.~{(\ref{eq:s12})}) and Eq.~{(\ref{eq:s14-1})} (Eq.~{(\ref{eq:s14-2})}) are consistent.

(iii)
\begin{equation}
\label{eq:s15}
\begin{aligned}
n_{C D, 3}(\tau)=&-\frac{\eta_{C}\eta_{D}}{4}\int_{T_{1}}^{T_{2}} dt \int_{-\Delta \tau_{c 0}/ 2}^{\Delta \tau_{c 0} / 2} d\tau_{c}\left\langle\left.\psi_{1}\right|_{A}\left\langle\left.\psi_{2}\right|_{B} a_{A}^{\dagger}(t) a_{A}^{\dagger}\left(t+\tau_{c}\right) a_{B}\left(t+\tau_{c}\right) a_{B}(t) \mid \psi_{1}\right\rangle_{A} \mid \psi_{2}\right\rangle_{B} \\
=&-\frac{\eta_{C} \eta_{D}}{4} \int_{T_{1}}^{T_{2}} d t \int_{-\Delta \tau_{c 0} / 2}^{\Delta \tau_{c 0} / 2} d \tau_{c} \alpha_{A 1}^{*}(t) \alpha_{A 1}^{*}\left(t+\tau_{c}\right) \alpha_{B 2}\left(t+\tau_{c}\right) \alpha_{B 2}(t) \\
=&-\frac{\eta_{C} \eta_{D}}{4}(2 \pi)^{-2} e^{-i 2 \omega_{0} \tau} \int_{T_{1}}^{T_{2}} d t \int_{-\Delta \tau_{c 0} / 2}^{\Delta \tau_{c 0} / 2} d \tau_{c} \int d \Omega_{A} \int d \Omega_{A}^{\prime} \int d \Omega_{B} \int d \Omega_{B}^{\prime} \\
&\times \alpha_{A}^{*}\left(\Omega_{A}\right) \alpha_{A}^{*}\left(\Omega_{A}^{\prime}\right) \alpha_{B}\left(\Omega_{B}\right) \alpha_{B}\left(\Omega_{B}^{\prime}\right) \\
&\times e^{i\left(\Omega_{A}+\Omega_{A}^{\prime}-\Omega_{B}-\Omega_{B}^{\prime}\right) t+i\left(\Omega_{A}^{\prime}-\Omega_{B}\right) \tau_{c}-i\left[\beta_{A}\left(\Omega_{A}\right) L_{A}+\beta_{A}\left(\Omega_{A}^{\prime}\right) L_{A}-\beta_{B}\left(\Omega_{B}\right) L_{B}-\beta_{A}\left(\Omega_{B}^{\prime}\right) L_{A}\right]} \\
=&-\frac{\eta_{C} \eta_{D}}{4} e^{-i 2 \omega_{0} \tau}\left\{\int d \Omega_{A} \alpha_{A}^{*}\left(\Omega_{A}\right) \alpha_{B}\left(\Omega_{A}\right) e^{-i\left[\beta_{A}\left(\Omega_{A}\right) L_{A}-\beta_{B}\left(\Omega_{A}\right) L_{B}\right]}\right\}^{2}.
\end{aligned}
\end{equation}

(iv)
\begin{equation}
\label{eq:s16}
\begin{aligned}
n_{C D, 4}(\tau)=&-\frac{\eta_{C} \eta_{D}}{4} \int_{T_{1}}^{T_{2}} d t \int_{-\Delta \tau_{c 0} / 2}^{\Delta \tau_{c 0} / 2} d \tau_{c}\left\langle\left.\psi_{1}\right|_{A}\left\langle\left.\psi_{2}\right|_{B} a_{B}^{\dagger}(t) a_{B}^{\dagger}\left(t+\tau_{c}\right) a_{A}\left(t+\tau_{c}\right) a_{A}(t) \mid \psi_{1}\right\rangle_{A} \mid \psi_{2}\right\rangle_{B} \\
=&-\frac{\eta_{C} \eta_{D}}{4} \int_{T_{1}}^{T_{2}} d t \int_{-\Delta \tau_{c 0} / 2}^{\Delta \tau_{c 0} / 2} d \tau_{c} \alpha_{B 2}^{*}(t) \alpha_{B 2}^{*}\left(t+\tau_{c}\right) \alpha_{A 1}\left(t+\tau_{c}\right) \alpha_{A 1}(t) \\
=&-\frac{\eta_{C} \eta_{D}}{4}(2 \pi)^{-2} e^{-i 2 \omega_{0} \tau} \int_{T_{1}}^{T_{2}} d t \int_{-\Delta \tau_{c 0} / 2}^{\Delta \tau_{c 0} / 2} d \tau_{c} \int d \Omega_{A} \int d \Omega_{A}^{\prime} \int d \Omega_{B} \int d \Omega_{B}^{\prime} \\
&\times \alpha_{B}^{*}\left(\Omega_{B}\right) \alpha_{B}^{*}\left(\Omega_{B}^{\prime}\right) \alpha_{A}\left(\Omega_{A}\right) \alpha_{A}\left(\Omega_{A}^{\prime}\right) \\
& \times e^{i\left(\Omega_{B}+\Omega_{B}^{\prime}-\Omega_{A}-\Omega_{A}^{\prime}\right) t+i\left(\Omega_{B}^{\prime}-\Omega_{A}\right) \tau_{c}\left[\beta_{B}\left(\Omega_{B}\right) L_{B}+\beta_{B}\left(\Omega_{B}^{\prime}\right) L_{B}-\beta_{A}\left(\Omega_{A}\right) L_{A}-\beta_{A}\left(\Omega_{A}^{\prime}\right) L_{A}\right]} \\
=&-\frac{\eta_{C} \eta_{D}}{4} e^{i 2 \omega_{0} \tau}\left\{\int d \Omega_{A} \alpha_{B}^{*}\left(\Omega_{A}\right) \alpha_{A}\left(\Omega_{A}\right) e^{-i\left[\beta_{B}\left(\Omega_{A}\right) L_{B}-\beta_{A}\left(\Omega_{A}\right) L_{A}\right]}\right\}^{2}.
\end{aligned}
\end{equation}

It is obvious that the terms $n_{C D, 3}(\tau)$ and $ n_{C D, 4}(\tau)$ oscillate quickly with variable $\tau$, and their sum is
\begin{equation}
\label{eq:s17}
n_{C D, 3}(\tau)+n_{C D, 4}(\tau)=-\frac{\eta_{C} \eta_{D}}{2} \cos \left(2 \omega_{0} \tau\right) \operatorname{Re}\left\{\int d \Omega_{A} \alpha_{B}^{*}\left(\Omega_{A}\right) \alpha_{A}\left(\Omega_{A}\right) e^{-i\left[\beta_{B}\left(\Omega_{A}\right) L_{B}-\beta_{A}\left(\Omega_{A}\right) L_{A}\right]}\right\}^{2}.
\end{equation}

The oscillating period of the term  $\cos \left(2 \omega_{0} \tau\right)$ is in the level of ~fs for visible and infrared light.  In experimental system, especially the one with optical fiber devices, the fluctuation of optical depth make $\tau$ fluctuate quickly with respect to the time scale of $T_{2}-T_{1}$  and therefore $n_{C D, 3}(\tau)+ n_{C D, 4}(\tau)$ in Eq.~{(\ref{eq:s17})} turns to be zero.

(v)
\begin{equation}
\label{eq:18}
\begin{aligned}
n_{C D, 5}(\tau)=&\frac{\eta_{C} \eta_{D}}{4} \int_{T_{1}}^{T_{2}} d t \int_{-\Delta \tau_{c 0} / 2}^{\Delta \tau_{c 0} / 2} d \tau_{c}\left\langle\left.\psi_{1}\right|_{A}\left\langle\left.\psi_{2}\right|_{B} a_{A}^{\dagger}(t) a_{B}^{\dagger}\left(t+\tau_{c}\right) a_{B}\left(t+\tau_{c}\right) a_{A}(t) \mid \psi_{1}\right\rangle_{A} \mid \psi_{2}\right\rangle_{B} \\
=&\frac{\eta_{C} \eta_{D}}{4} \int_{T_{1}}^{T_{2}} d t \int_{-\Delta \tau_{c 0} / 2}^{\Delta \tau_{c 0} / 2} d \tau_{c}\left|\alpha_{A 1}(t)\right|^{2}\left|\alpha_{B 2}\left(t+\tau_{c}\right)\right|^{2} \approx\left|\alpha_{A0}\right|^{2}\left|\alpha_{B0}\right|^{2}.
\end{aligned}
\end{equation}

(vi)
\begin{equation}
\label{eq:s19}
\begin{aligned}
n_{C D, 6}(\tau)=&\frac{\eta_{C} \eta_{D}}{4} \int_{T_{1}}^{T_{2}} d t \int_{-\Delta \tau_{c 0} / 2}^{\Delta \tau_{c 0} / 2} d \tau_{c}\left\langle\left.\psi_{1}\right|_{A}\left\langle\left.\psi_{2}\right|_{B} a_{B}^{\dagger}(t) a_{A}^{\dagger}\left(t+\tau_{c}\right) a_{A}\left(t+\tau_{c}\right) a_{B}(t) \mid \psi_{1}\right\rangle_{A} \mid \psi_{2}\right\rangle_{B} \\
=&\frac{\eta_{C} \eta_{D}}{4} \int_{T_{1}}^{T_{2}} d t \int_{-\Delta \tau_{c 0} / 2}^{\Delta \tau_{c 0} / 2} d \tau_{c}\left|\alpha_{A 1}\left(t+\tau_{c}\right)\right|^{2}\left|\alpha_{B 2}(t)\right|^{2} \approx\left|\alpha_{A0}\right|^{2}\left|\alpha_{B0}\right|^{2}.
\end{aligned}
\end{equation}

At the last step in Eq.~{(\ref{eq:s12})} , the assumption that $ T_{2}- T_{1}$ and $\Delta \tau_{c 0}$ are remarkably larger than the temporal width of the dispersion-broadened wave-packets is still used.

(vii)
\begin{equation}
\label{eq:s20}
\begin{aligned}
n_{C D, 7}(\tau)=&-\frac{\eta_{C} \eta_{D}}{4} \int_{T_{1}}^{T_{2}} d t \int_{-\Delta \tau_{c 0} / 2}^{\Delta \tau_{c 0} / 2} d \tau_{c}\left\langle\left.\psi_{1}\right|_{A}\left\langle\left.\psi_{2}\right|_{B} a_{A}^{\dagger}(t) a_{B}^{\dagger}\left(t+\tau_{c}\right) a_{A}\left(t+\tau_{c}\right) a_{B}(t) \mid \psi_{1}\right\rangle_{A} \mid \psi_{2}\right\rangle_{B} \\
=&-\frac{\eta_{C} \eta_{D}}{4}(2 \pi)^{-2} \int_{T_{1}}^{T_{2}} d t \int_{-\Delta \tau_{c 0} / 2}^{\Delta \tau_{c 0} / 2} d \tau_{c} \alpha_{A 1}^{*}(t) \alpha_{B 2}^{*}\left(t+\tau_{c}\right) \alpha_{A 1}\left(t+\tau_{c}\right) \alpha_{B 2}(t) \\
=&-\frac{\eta_{C} \eta_{D}}{4}(2 \pi)^{-2} \int_{T_{1}}^{T_{2}} d t \int_{-\Delta \tau_{c 0} / 2}^{\Delta \tau_{c 0} / 2} d \tau_{c} \int d \Omega_{A} \int d \Omega_{A}^{\prime} \int d \Omega_{B} \int d \Omega_{B}^{\prime} \alpha_{A}^{*}\left(\Omega_{A}\right) \alpha_{B}^{*}\left(\Omega_{B}\right) \alpha_{A}\left(\Omega_{A}^{\prime}\right) \alpha_{B}\left(\Omega_{B}^{\prime}\right) \\
&\times e^{i\left(\Omega_{A}-\Omega_{B}^{\prime}+\Omega_{B}-\Omega_{A}^{\prime}\right) t+i\left(\Omega_{B}-\Omega_{A}^{\prime}\right) \tau_{c}+i\left(\Omega_{B}-\Omega_{B}^{\prime}\right) \tau-i\left[\beta_{B}\left(\Omega_{B}^{\prime}\right) L_{B}-\beta_{B}\left(\Omega_{B}\right) L_{B}+\beta_{A}\left(\Omega_{A}^{\prime}\right) L_{A}-\beta_{A}\left(\Omega_{A}\right) L_{A}\right]} \\
=&-\frac{\eta_{C} \eta_{D}}{4}\left|\int d \Omega \alpha_{A}^{*}(\Omega) \alpha_{B}(\Omega) e^{-i \Omega \tau+i\left[\beta_{A}(\Omega) L_{A}-\beta_{B}(\Omega) L_{B}\right]}\right|^{2}.
\end{aligned}
\end{equation}

(viii)
\begin{equation}
\label{eq:s21}
\begin{aligned}
n_{C D, 8}(\tau)=&-\frac{\eta_{C} \eta_{D}}{4} \int_{T_{1}}^{T_{2}} d t \int_{-\Delta \tau_{c 0} / 2}^{\Delta \tau_{c 0} / 2} d \tau_{c}\left\langle\left.\psi_{1}\right|_{A}\left\langle\left.\psi_{2}\right|_{B} a_{B}^{\dagger}(t) a_{A}^{\dagger}\left(t+\tau_{c}\right) a_{B}\left(t+\tau_{c}\right) a_{A}(t) \mid \psi_{1}\right\rangle_{A} \mid \psi_{2}\right\rangle_{B} \\
=&-\frac{\eta_{C} \eta_{D}\left|\alpha_{A}\right|^{2}\left|\alpha_{B}\right|^{2}}{4}(2 \pi)^{-2} \int_{T_{1}}^{T_{2}} d t \int_{-\Delta \tau_{c 0} / 2}^{\Delta \tau_{c 0} / 2} d \tau_{c} \alpha_{B 2}^{*}(t) \alpha_{A 1}^{*}\left(t+\tau_{c}\right) \alpha_{B 2}\left(t+\tau_{c}\right) \alpha_{A 1}(t) \\
=&-\frac{\eta_{C} \eta_{D}}{4}\left|\int d \Omega \alpha_{A}^{*}(\Omega) \alpha_{B}(\Omega) e^{-i \Omega \tau+i\left[\beta_{A}(\Omega) L_{A}-\beta_{B}(\Omega) L_{B}\right]}\right|^{2}.
\end{aligned}
\end{equation}
After summing Eqs.~{(\ref{eq:s11})}, {(\ref{eq:s12})}, {(\ref{eq:s15})}-{(\ref{eq:s21})} up, we can finally get
\begin{equation}
\label{eq:s22}
\begin{aligned}
n\left(\tau\right)=&\frac{\eta_{C} \eta_{D}\left|\alpha_{A}\right|^{2}\left|\alpha_{B}\right|^{2}}{4}\left\{\frac{\left|\alpha_{A}\right|^{2}}{\left|\alpha_{B}\right|^{2}}+\frac{\left|\alpha_{B}\right|^{2}}{\left|\alpha_{A}\right|^{2}}+2-\right. \\
&\left.\frac{2}{\left|\alpha_{A}\right|^{2}\left|\alpha_{B}\right|^{2}}\left|\int d \Omega \alpha_{A}^{*}(\Omega) \alpha_{B}(\Omega) e^{-i\left[\Omega \tau+\beta_{A}(\Omega) L_{A}-\beta_{B}(\Omega) L_{B}\right]}\right|^{2}\right\}.
\end{aligned}
\end{equation}

Now, we introduce some assumptions for conveniently comparing Eq.~{(\ref{eq:s22})} to the experiments. Firstly, the averaged photon number in the two single-photon wave-packets are both 1 i.e., $\left|\alpha_{A0}\right|^{2}=\left|\alpha_{B0}\right|^{2}=1$. For convenience, $\alpha_{A0}$ and $\alpha_{B0}$ are treated as 1. Secondly, the single-photon wave-packets before passing through the dispersive medium are indistinguishable, and their temporal profile is Gaussian, i.e.,
\begin{equation}
\label{eq:s23}
\alpha_{A, B}(t)=\left(T_{0} \sqrt{\pi}\right)^{-1 / 2} e^{-\frac{t^{2}}{2 T_{0}^{2}}}.
\end{equation}
Thirdly, the effect of higher order dispersions is negligible, and $\beta_{A, B}(\Omega)$ can be approximated by truncating the Taylor expansion to the second order, i.e.,
\begin{equation}
\label{eq:s24}
\beta_{A, B}(\Omega) \approx \beta_{A, B}(0)+\beta_{A, B}^{(1)}(0) \Omega+\beta_{A, B}^{(2)}(0) \Omega^{2},
\end{equation}
where $\beta_{A, B}^{(1)}(0)$ and $\beta_{A, B}^{(2)}(0)$ are the first- and second-order dispersion coefficients, respectively.

Under the three assumptions above, Eq.~{(\ref{eq:s22})} gets
\begin{equation}
\label{eq:s25}
n\left(\tau\right)=\eta_{C} \eta_{D}\left(1-\frac{T_{0}^{2}}{\sqrt{4 T_{0}^{2}+\alpha^{2}}} e^{-2\left(\tau-\delta \tau\right)^{2} /\left[4 T_{0}^{2}+\left(\alpha / T_{0}\right)^{2}\right]}\right),
\end{equation}
where $\delta \tau=\beta_{A}^{(1)}(0) L_{A}-\beta_{B}^{(1)}(0) L_{B}$ and $\alpha=\beta_{A}^{(2)}(0) L_{A}-\beta_{B}^{(2)}(0) L_{B}$. It is obvious that the coincidence count per trial $P(\tau)$ can be obtained by normalizing $n\left(\tau\right)$ in Eq.~{(\ref{eq:s25})}, i.e.,
\begin{equation}
\label{eq:s26}
P(\tau)=1-\frac{T_{0}^{2}}{\sqrt{4 T_{0}^{2}+\alpha^{2}}} e^{-2(\tau-\delta \tau)^{2}/\left[4 T_{0}^{2}+\left(\alpha / T_{0}\right)^{2}\right].}
\end{equation}

In Eq.~{(\ref{eq:s26})}, the $P(\tau)$ versus $\tau$ forms the HOM interference curve and shows a dip centring at $\tau=\delta \tau$. The full width at half maximum (FWHM) of this dip is
\begin{equation}
\label{eq:s27}
d=2 \sqrt{\ln 4} \sqrt{T_{0}^{2}+\left(\alpha/{2 T_{0}}\right)^{2}}.
\end{equation}

When the two wave-packets experience identical dispersion, i.e., $\delta \tau =0$ and $\alpha =0$, Eq.~{(\ref{eq:s15})} degrades to
\begin{equation}
\label{eq:s28}
P(\tau)=1-\frac{1}{2} e^{-\tau^{2} /\left(2 T_{0}^{2}\right)}
\end{equation}
and its FWHM is
\begin{equation}
\label{eq:s29}
d=2 \sqrt{\ln 4}{T_{0}}=\sqrt{2}{T_{FWHM}},
\end{equation}
where $T_{FWHM}$ is the full width at half maximum of the wave-packets. If we let  $\beta_{A, B}(\Omega)=0$, i. e, there does not exist dispersion along the propagation pathways of the two wave-packets, we can get a expression of $P(\tau)$ identical to Eq.~{(\ref{eq:s28})}. This indicates that the HOM interference curve will not be affected by the balanced dispersion experienced by the two wave-packets. Therefore, we can obtain the original width of single photon wave-packet as the $1/\sqrt{2}$ times the HOM interference curve even though the wave-packets have been broadened significantly by the dispersive manipulation. According to Eq.~{(\ref{eq:s27})}, when the dispersive manipulation in two paths are unbalanced, the difference of the second-order dispersion effect between two paths can accurately mapped to the HOM interference curve by
\begin{equation}
\label{eq:s30}
\alpha=T_{0} \sqrt{\left(d^{2}/2\ln 2-4 T_{0}^{2}\right)},
\end{equation}
which can be utilized to measure the second-order dispersion effect of an unknown optical material.

\section{\label{app:2}Methods}
\begin{figure}
    \centering
    \includegraphics[width=17.2 cm]{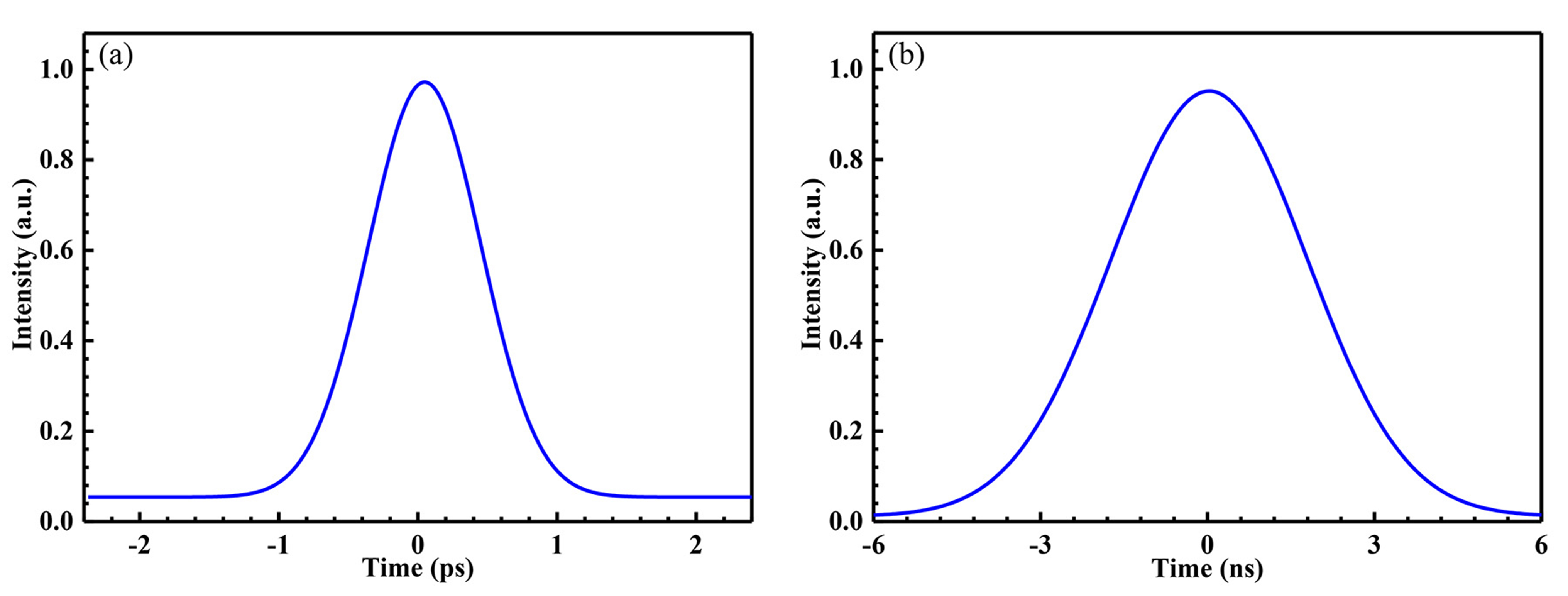}
    \caption{(a) Output of a mode-locked pulse laser, measured by a second-order autocorrelator. (b) Output of a mode-locked pulse laser after the dispersive manipulation, recorded by the single-photon detection and a TDC.}
    \label{fig:figa1}
\end{figure}

The pulsed light source is selected as a passively mode-locked fiber laser with the central wavelength of 1565 nm, the pulse duration of $0.798\pm0.010$ ps by second-order autocorrelation method, the repetition rate of 5 MHz and the 3-dB spectral bandwidth of 4.3 nm as shown in Fig.~{\ref{fig:figa1}}(a) . The laser pulses are attenuated to the single-photon level with the mean photon number of 0.03 per wave-packet through a variable optical attenuator (VOA1). To simulate the two single-photon wave-packets, a 50:50 single-mode fiber coupler (SMC) is used to separate the attenuated laser pulses into two paths, resulting in the average photon number of 0.015 per wave-packet in each path. Two wave-packets are injected into a HOM interferometer, which consists of two variable optical attenuators (VOA2 and VOA3), two polarization beam splitters (PBS1 and PBS2), a 50:50 polarization-maintaining fiber coupler (PMC), and a fiber pigtailed variable optical delay line (Delay). VOA2 and VOA3 in two arms are used to adjust the mean photon number in the two optical paths, which ensure that the mean photon numbers of two paths are the same. PBS1 and PBS2 are used to ensure the indistinguishability in polarization. The relative time delay is introduced by an optical delay-line (MDL-002, General Photonics) with the accuracy of 10 fs and a maximum range of 560 ps. Two output ports of HOM interferometer are connected with two superconducting nanowire single photon detectors (SNSPDs) with a detection efficiency of the 68$\%$. The electronic signals generated by photon detection events are input into a time to digital converter (TDC, ID 900, ID Quantique) to obtain the coincidence counts. The HOM-dip can be observed in the HOM interference curve, i.e. coincidence counts versus the relative time delay between two paths.

In the experiment, to demonstrate that the intrinsic width of a mode-locked laser can be measured after the dispersive manipulation, we connect the output port of a mode-locked laser with a 50 km-long fiber spool, and attenuate the laser pulses into the single-photon level with VOA1, VOA2 and VOA3, then send the attenuated laser pulses into HOM interferometer.  In our demonstration, the widths of laser pulses after the dispersive manipulation are also measured through the single-photon detection and a TDC. The measured results are shown in Fig.~{\ref{fig:figa1}} (b). It shows that after propagating along a 50 km-long fiber, the width of pulse laser has been broadened to 4.1 ns including a time jitter of about 270 ps in the single-photon detector and TDC. To demonstrate that the second-order dispersion coefficient of an unknown optical medium can be measured through HOM interference, we first attenuate the mode-locked laser pulses and then separate them into two paths. In one of the two paths we connect a piece of 80 meter-long fiber as the dispersive optical medium for the test. These connections are switched in the experiment.

